\documentclass{PoS}

\usepackage{latexsym}
\usepackage{graphics}
\usepackage{epsfig}
\usepackage{overpic}

\usepackage{amsmath} %, amssymb, amsthm}
\def\pt{p_{\rm T}}
\def\av#1{\langle #1 \rangle}

\title{Constraints on string percolation model from anomalous centrality evolution data in Au-Au collisions at $\mathbf{\sqrt{s_{NN}}=}$ 62 and 200 GeV}

\ShortTitle{Constraints on string percolation model}

\author{\speaker{Grigory Feofilov}%
\\
 St.Petersburg State University\\
 E-mail: \email{feofilov@hiex.phys.spbu.ru}}

\author{Igor Altsybeev \\
 St.Petersburg State University\\
}

\author{Olga Kochebina \\
 Laboratoire de l'Accelerateur Lineaire (LAL), CNRS : UMR8607 - IN2P3 - Universite Paris XI - Paris Sud\\
}

\abstract{Anomalous centrality evolution of two-particle angular correlations observed in Au-Au collisions at $\sqrt{s_{NN}} = 62$ and 200 GeV and the onset of ridge structures are considered in the model of interacting quark-gluon strings.
We assume that at the given energy of nucleus-nucleus collisions
the critical energy density may be reached at the specific centrality.
In a string percolation model
this might be treated 
equivalently to a formation of a large cluster of strings characterized by the critical string density,
with a size comparable to the whole area of interaction of two nuclei.
This hypothesis allows to define some constraints on the string percolation model using data on transitional centralities in Au-Au collisions at these two energies.
Results are extrapolated to the LHC energy where high string densities (exceeding the critical value) are confirmed for all classes of centralities in Pb-Pb collisions.

Interaction between strings inside  large clusters formed in nucleus-nucleus collisions is considered in a simplified Monte Carlo model.
This model  is applied to the qualitative analysis of the onset of collectivity and the ridge formation in Pb-Pb collisions. It is shown that the approach of the repulsive string-string interaction is capable to explain the appearance of elliptic and triangular flow observed in nucleus-nucleus collisions at RHIC and LHC energies. 
 }

\FullConference{XXII International Baldin Seminar on High Energy Physics Problems \\
 15-20 September, 2014\\
 JINR, Dubna, Russia}

\begin{document}

\maketitle

%%%%%%%%%%%%%%%%%%%%%%%%%%%%%%%%%%%%%%%%%%%%
%% MAINMATTER
%%%%%%%%%%%%%%%%%%%%%%%%%%%%%%%%%%%%%%%%%%%%

\section{Introduction}
\label{intro}

The first predictions of the azimuthal asymmetry of multiple-production of secondary hadrons
in high energy nucleus-nucleus collisions were done \cite{Abramovsky-1980, Abramovsky-1988} %on the base of
using the concept of interacting color flux tubes (strings).
The color strings may be viewed as tubes of the color field created by the colliding partons~\cite{Capella,Kaidalov,Armesto2000}.
Production of particles goes via spontaneous emission of quark-antiquark pairs in this color field. These strings are the phenomenological objects extended in rapidity and are related to the cut Pomerons. Their cross-section in the transverse plane is considered as small discs of $\pi r_{0}^2$ area, where $r_{0}$ is the string radius usually taken to be about 0.2~fm. With growing energy and/or atomic number of colliding particles the number of strings grows, therefore they start to overlap and may interact.
In case of existence of string-string interaction, the event-by-event fluctuations of the initial geometry of  collisions should manifest themselves as the azimuthal ($\phi$) anisotropy in two-particle correlations functions. The second important outcome of this hypothesis \cite{Abramovsky-1988} is that
this spatial $\phi$ asymmetry will be also manifested as
 the long-range (extended in pseudorapidity $\eta$) correlations.

Experimental evidences of the long-range azimuthal anisotropy in two-particle correlations in heavy-ion collisions at RHIC and LHC are well-known.
The ridge was defined as a two-particle correlation structure
relatively narrow in azimuthal angle and extended over
several units in pseudo-rapidity \cite{STAR 2006-05,Putschke,STAR
2006}.
These structures were also observed both in $\pt$-integrated and in
special $\pt$-selected analyses
of the 62 and 200 GeV Au-Au and Cu-Cu
data (one may see a detailed overview of STAR ridges in a recent work
\cite{star-ridges}).
Recently the experimental ridge landscape was considerably broadened by
the observation of the CMS Collaboration at the LHC when the unexpected long-range
azimuthal two-particle correlations where found in pp
collisions~\cite{CMS-ridge-pp}.
Ridge structures were also reported in Pb-Pb and in p-Pb collisions at LHC
\cite{CMS-ridge-PbPb, CMS-ridge-pPb, ALICE-ridge-PbPb-harmonics}.

The onset of the ridge and the role of initial conditions in the ridge formation
were considered by a rather large number of theoretical models that were motivated by the experimental discoveries.
Several models were proposed to explain qualitatively an
origin of the ridge  using various concepts
like an interaction of high-$\pt$ partons or jets with medium,
formation of jets in small-$\pt$, parton-jets collisions, glasma flux tubes with
radial flow etc. (see references in~\cite{Putschke,star-ridges,CMS-ridge-PbPb}).
The Color Glass Condensate (CGC) model~\cite{ridge-cgc} in addition to the long-range rapidity
correlation points to the possibility of
intrinsic angular correlation which is assumed
to come from the particle production process due to glasma tubes formation on transverse distance scales 1/$Q_s$ much smaller than the proton size (here $Q_s$ is the saturation scale of the colliding nuclei).
 In \cite{Perc_ridge} the string percolation phenomenology was compared
 to CGC results on effective string or glasma flux tube intrinsic correlations, including the ridge
phenomena and long-range forward-backward azimuthal correlations. Color string
percolation model and its similarities with the CGC are discussed in \cite{Paj-2011}.
Fourier harmonics decomposition of the two-particle azimuthal %asymmetry in
correlations %function
 in nucleus-nucleus collisions was found to describe %contribute notably to the formation of
various ridge structures observed in the experiment \cite{ALICE-ridge-PbPb-harmonics, ALICE-H}.
 However, the dynamical origin of the harmonics %higher-order harmonics
and of the onset of these collective phenomena are still not clear enough. %understood. %Therefore,

In the present work we use %are using,
as the main working
hypothesis the interaction among the quark-gluon strings formed in the nucleus-nucleus collisions.
In Section~\ref{onset}, estimations of string density that might be reached in nucleus-nucleus collisions are done.
Following \cite{OK-2010}, we assume here that the intriguing phenomena of sharp change in two-particle correlation function, observed for the first time in Au-Au collisions at certain collision centralities at $\sqrt{s_{NN}}=62$ and 200 GeV~\cite{Daugherity}, is related to the critical string density reached in the interaction region.
In Section~\ref{MC-toy}, a toy-model with interacting strings in nucleus-nucleus collisions~\cite{TOY} is applied for the analysis of the topology of two-particle correlations to study the phenomena of the onset of the azimuthal peculiarities.
In this model,
a string repulsion is considered as the collective effect of a large number of interacting strings.
The Monte Carlo model allows to understand in a qualitative way the formation of the initial conditions representing the dynamic origin for the elliptic flow and for the higher-order components of the two-particle angular correlations  observed in nucleus-nucleus collisions.

\section{String density in nucleus-nucleus collisions}
\label{onset}

 In this section we use the string percolation model for the analysis of the onset of the long-range correlations
in Au-Au collisions at RHIC
and estimate string densities of nucleus-nucleus collisions at different collision energies and centralities.

\subsection{Onset of ridge phenomena in Au-Au collisions at RHIC}

The very first experimental data on the ridge onset were obtained in
detailed study of centrality dependence of two-particle correlations done
by STAR in Au-Au collisions at 62 and 200~GeV at RHIC for all charged
hadrons with rather low-$\pt$ ($\pt > 0.15$ GeV/{\it c})~\cite{Daugherity}.
These preliminary results revealed that the "soft ridge" structure appears in Au-Au
collisions after reaching certain collision centrality that might
be characterized by definite ("critical") number of participating
nucleons ($N_{part}^{crit}(\sqrt{s})$).
These "critical centralities"
were found to be different for two collision energies: the onset of the
ridge was observed in Au-Au at approximately 55\% centrality for collision energy
$\sqrt{s_{NN}}$ 200~GeV and at about 40\% for $\sqrt{s_{NN}}=62$~GeV~\cite{Daugherity}. One may see from the data~\cite{Daugherity} that at $\sqrt{s_{NN}}=62$~GeV this phenomenon starts at $N_{part}^{crit}\approx90$, while at $\sqrt{s_{NN}}=200$~GeV the relevant threshold is marked by $N_{part}^{crit}\approx40$.
The uncertainties of these values of $N_{part}^{crit}$, extracted from the RHIC data, produce some systematic error that is taken in account in our calculations.

Moreover, it was also found in~\cite{Daugherity} that transverse particle density
\begin{equation}
\label{rho}
\tilde{\rho}=\frac{3}{2}~{\frac{dN_{ch}}{dy}}/{\av{S}}
\end{equation}
brings the transition points for these two energies to the same value $2.6\pm0.2$~fm$^{-2}$.
Here $\frac{dN_{ch}}{dy}$ is the total charge particle multiplicity per rapidity unit at a given centrality, $\av{S}$ is the collision overlap area, the factor $\frac{3}{2}$ appears because both charge and neutral particles are taken into account. It is important to note that both $\frac{dN_{ch}}{dy}$ and $\av{S}$ in Eq.~\ref{rho} depend on centrality of collisions defined by the number of nucleons-participants $N_{part}$.

\begin{figure}
\centering
 \includegraphics[width=0.75\textwidth]{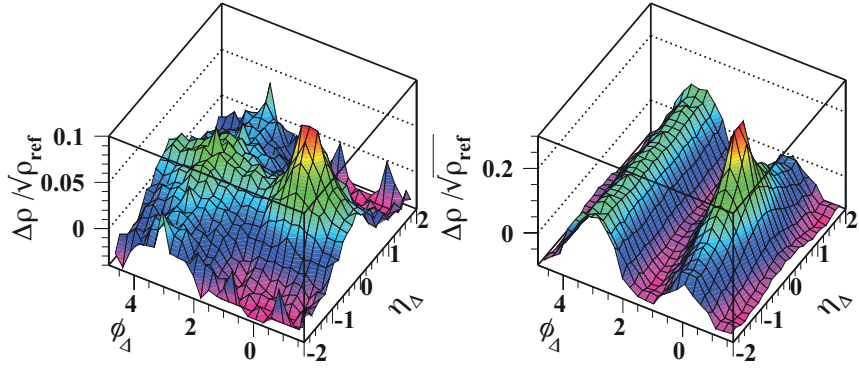}
\caption{Example of changes in 2D two-particle angular correlations with centrality of Au-Au collisions at $\sqrt{s}$ = 200 GeV: from peripheral
(84$\%$ - 93$\%$) to semi-peripheral (55$\%$ - 64$\%$)
collisions (see more detailed plots in \cite{STAR-2012}). }
\label{ris:onset}
\end{figure}

The detailed analysis of anomalous evolution of two-particle correlations with energy and centrality of Au-Au collisions was followed in \cite{STAR-2012}.
The sudden change in 2D angular correlations, 
observed by STAR
at some critical centrality (example is shown in Figure~\ref{ris:onset})
motivated our application of the
string percolation model to describe this phenomenon.

\subsection{String density in Au-Au collisions in string percolation model}
\label{density}

In the present study we assume that the onset of the low-$\pt$ manifestation of a near-side ridge phenomenon in Au-Au collisions discussed above,
is related to the critical quark-gluon string density reached at certain centrality. Under these conditions, a "macroscopic" cluster could appear, which would be composed of a large number of overlapped strings extended in rapidity. % and localized in azimuth.
Such a cluster might be considered as a new kind of source emitting correlated particles. Cluster formation and the azimuthal effects in correlation functions might be due to some process that starts to be visible above the percolation threshold.

To characterize mathematically the string density, a dimensionless percolation parameter $\tilde{\eta}$ is introduced~\cite{Armesto1996,Nardi,Braun2000}:

\begin{equation}
\label{perc}
\tilde{\eta}=\frac{\pi r_{0}^2 N_{str}}{\av{S}}\ .
\end{equation}

\noindent Here $\av{S}$ is the transverse area of the overlap of colliding nuclei,
$N_{str}$ is a number of strings.
The critical value of the parameter $\tilde{\eta}$ marking the percolation transition ($\tilde{\eta}^{crit}$) could be calculated from the geometrical considerations  
and is estimated to be $\tilde{\eta}^{crit}\approx~1.12-1.175$~\cite{BraunPRL}, string radius
is usually taken as $r_{0}=0.2-0.3$~fm~\cite{Armesto2000,Dias,Pajares}.
In our calculations we use $\tilde{\eta}^{crit}~\approx~1.15\pm0.03$ and $r_{0}~=~0.25$~fm. (We have to note here that only the product of ${r_{0}^2 N_{str}}$ could be constrained using Eq.~\ref{perc}. So that one will get different value of $N_{str}$ in case of using the different value of $r_{0}$).

The number of particle emitting strings $N_{str}$ generally depends on the centrality of nuclus-nucleus collision, on the type of colliding system and on the collision energy $\sqrt{s}$.
In our approach, $N_{str}$ and the overlap area $\av{S}$
depend on $N_{part}$. However, these variables could not be measured directly.
In this work we exclude $\av{S}$ from the estimations by considering the ratio
$\tilde{\rho}^{crit}/\tilde{\eta}^{crit}$ at the "critical" point, that marks the onset of the low-$\pt$ ridge manifestation mentioned above.
Thus at the critical point one may obtain the following value:

\begin{equation}
\label{ratio}
\frac{\tilde{\rho}^{crit}(N_{part})}{\tilde{\eta}^{crit}(N_{part})}=\frac{3}{2} \frac{1}{\pi r_{0}^2} \frac{dN_{ch}}{dy} \frac{1}{N_{str}} = 2.3 \pm 0.2 fm^{-2},
\end{equation}
here the error is coming mainly from the systematic uncertainties of $\tilde{\rho}^{crit}$ and $\tilde{\eta}^{crit}$.
The total number of  strings $N_{str}$ at the "critical" points in Au-Au collisions at $\sqrt{s_{NN}}=62$ and 200 GeV could be easily found
from the Eq.~\ref{ratio}.
The results of the calculations are presented in Table~\ref{tab:results-AA}.

In order to make rough estimates for the dependence of the mean number of strings formed in nucleus-nucleus collisions vs. energy and centrality,
we use
the concept of valence and sea strings. 
A number of the valence strings $N_{V}$ is defined by $N_{part}$
and a number of the sea strings $N_{S}$ is proportional to~$N_{coll}$,
with a coefficient $a$.
For the total number of the strings $N_{str}$, formed in nucleus-nucleus collisions at some given energy, we use the following parametrization:
\begin{equation}
\label{a-prop}
N_{str}=N_{V}+N_{S}=N_{part}+a\cdot N_{coll}.
\end{equation}
The number $N_{str}$ can be estimated using Eq.~\ref{ratio} at the "critical" points, characterized by certain %measured
values of $N_{part}$ and estimated $N_{coll}$, after that the  coefficient  $a$ can calculated from Eq.~\ref{a-prop}.
Results of the calculations of the parameter $a$ for $\sqrt{s_{NN}}$ = 62 and 200~GeV are presented in Table~\ref{tab:results-AA}.
In the third line of the Table we also added results of our previous similar estimations~\cite{j-psi} in the framework of string percolation model based on the observed threshold of anomalous suppression of $J/\psi$ in Pb-Pb collisions at $\sqrt{s_{NN}}= 17.3$~GeV at SPS.

\begin{table}[h]
\caption{Number of participants, density of charged particles at midrapidity per pair of participants, the total number of strings, the number of sea strings and nucleon collisions parameter $a$ obtained at the "critical" points of Au-Au collisions at $\sqrt{s_{NN}}=62$~GeV and 200~GeV and for Pb-Pb collisions at $\sqrt{s_{NN}}=17.3$~GeV. The calculations are done for string radius $r_{0}=0.25$~fm.}
\label{tab:results-AA}
\begin{tabular}{lllllll}
\hline\noalign{\smallskip}
$\sqrt{s},~GeV~~~~~~~$ & $N_{part}$~~& $(dN_{ch}/d\eta)$/$(0.5*N_{part})$~~& $N_{str}$~~~~~~~&
 $N_{s}$~~~~~~~~& $N_{coll}$~~~~~~~& $a$~~~~~~~~\\
\noalign{\smallskip}\hline\noalign{\smallskip}
200(Au-Au) & 40 & $2.97\pm0.30$~\cite{PHOBOS130GeV-200GeV} & $194\pm25$ & $155\pm23$& $59\pm4$ & $2.6\pm0.4$ \\
62 (Au-Au)& 90 & $2.30\pm0.23$~\cite{PHOBOS62} & $352\pm28$ & $262\pm23$ & $167\pm4$ & $1.6\pm0.2$\\
17.3 (Pb-Pb)& 110 & $1.62\pm0.21$~\cite{j-psi-pb} & $302\pm45$ & $192\pm30$ & $158\pm5$& $1.2\pm0.2$ \\
\noalign{\smallskip}\hline
\end{tabular}
\vspace*{1cm}
\end{table}

\begin{figure}
%\begin{center}
\centering
$\begin{array}{ccc}
\begin{overpic}[width=0.62\textwidth, clip=true, trim=10 0 10 10]
{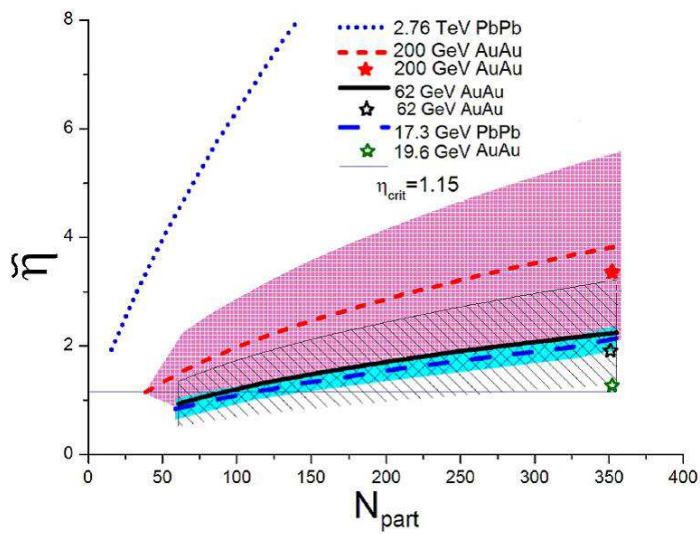}
\end{overpic}
\end{array}$
%\end{center}
\caption{
Centrality dependence of percolation parameter $\tilde{\eta}$ in nucleus collisions at various energies. Shaded areas are representing the uncertainties of the calculations (uncertainties for $\sqrt{s}$=~2.76~TeV are not shown).
Markers represent independent estimates~\cite{nucleonika} of $\tilde{\eta}$ in very central collisions. See details in \cite{OK-2010}.
}
\label{fig:aAndEta}
\end{figure}

It is possible to extrapolate the parameter $a$ to other energies and centralities of collisions with definite uncertainties (see details in \cite{OK-2010}).
In order to get the centrality dependence of $\tilde{\eta}$ in nucleus-nucleus collisions, the calculations of the interaction area $\av{S}$ are performed by applying the relation $\av{S}\sim~N_{part}^{2/3}$~\cite{S_Npart}. The coefficient of proportionality is derived from the information obtained at the "critical" point of the transverse particle density $\tilde{\rho}$ as it is mentioned above in Eq.~\ref{rho}. The Modified Glauber model~\cite{Ivanov} is used here for calculations of $N_{part}$ and $N_{coll}$.

One may see on Figure~\ref{fig:aAndEta} that rather large values of average string density $\tilde{\eta}$ exceeding considerably the "critical" density value are obtained.
The observation of the ridge at SPS energies, reported in \cite{na49-ridge}, could be the first experimental hint, confirming that the critical string density is reached in central Pb-Pb collisions at $\sqrt{s} = 17.3$ GeV. At the same time string density acquired in Pb-Pb collisions at the LHC energies exceeds the percolation threshold in all centrality classes.

\section{String-string interaction in Monte Carlo toy model }
\label{MC-toy}

 \subsection{Monte Carlo toy model}

Interaction between color strings formed in nucleus-nucleus collisions might produce clear experimental manifestations in two-particle angular correlations.
An exact form of this string-string interactions is not known.
As it was proposed in \cite{Abramovsky-1988}, it could be attractive or repulsive depending on the directions of the chromo-electric field inside the string.
The issue of the string-string interaction has not yet been systematically addressed till recently. One can find the overview
of the problem in~\cite{strings-interaction}. The magnitude of this interaction in string tension units was found to be small ($\sim 10^{-1}-10^{-2}$~\cite{strings-interaction}).
 It is natural that deeper basic fundamental understanding of string-string interaction is  required.  "Mesonic clouds" around the color flux tubes and exchange of the scalar $\sigma$-meson 
were  proposed in~\cite{strings-interaction} and  could be considered as the origin of interaction.
\vspace*{0.2cm}
\begin{figure}[h]
%\begin{center}
\centering
$\begin{array}{ccc}
\begin{overpic}[width=0.38\textwidth, clip=true, trim=0 0 20 0]
{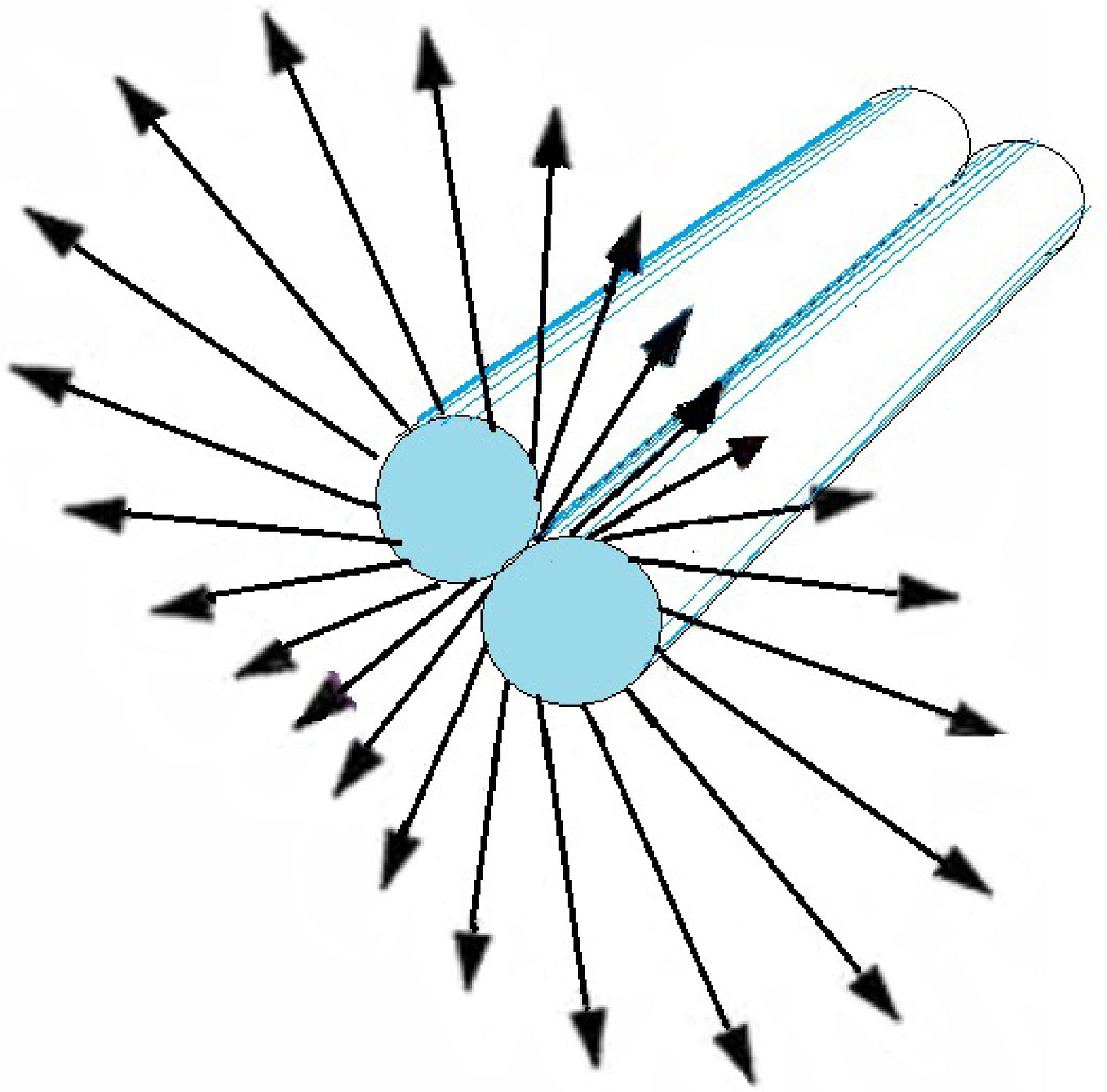}
\end{overpic}
&
\begin{overpic}[width=0.38\textwidth, clip=true, trim=0 0 0 0]
{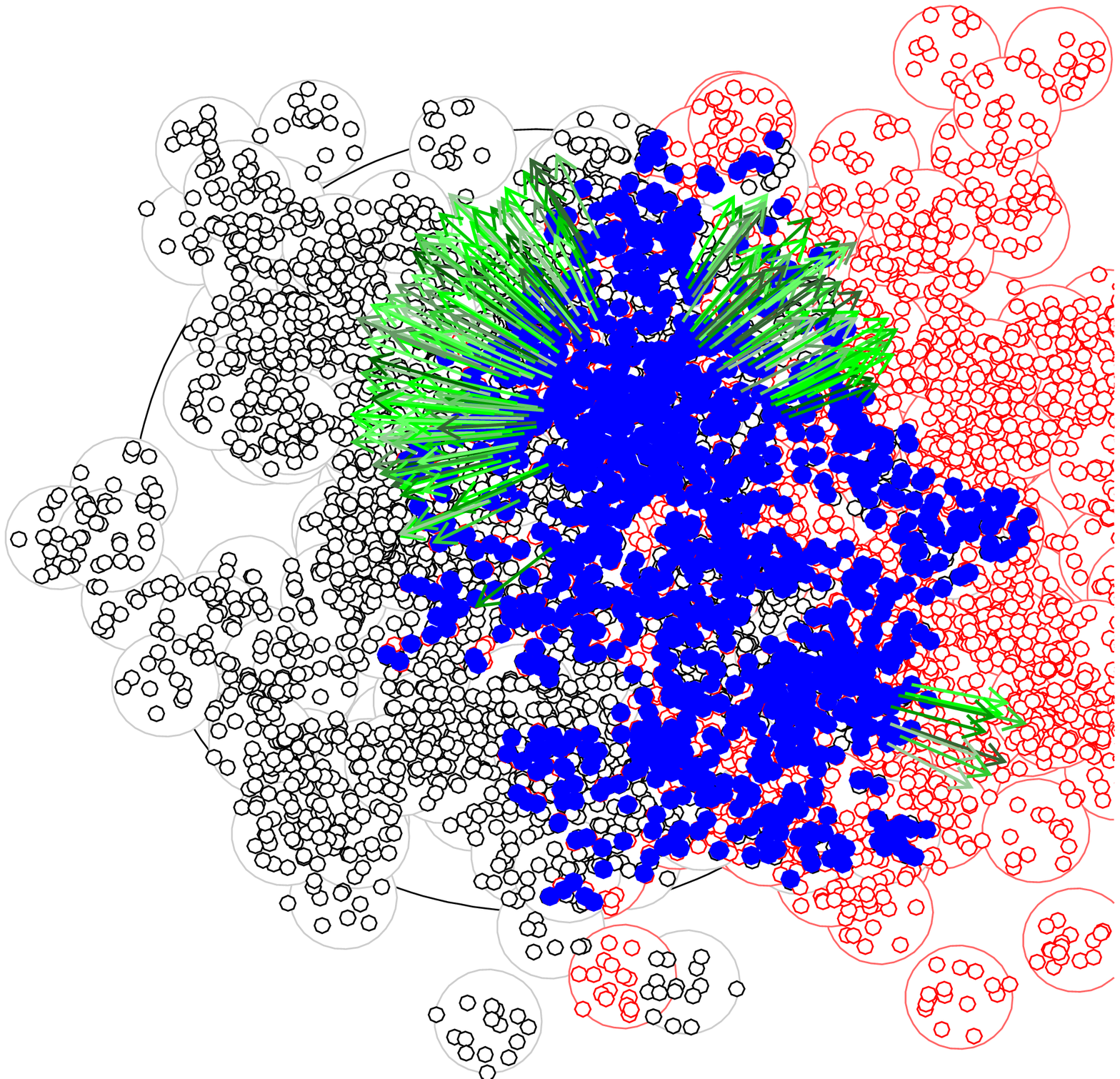}
\end{overpic}
\end{array}$
%\end{center}
\caption{
Left: Illustration  of azimuthally asymmetric flow in the transverse plane generated by two repulsive color flux tubes boosted in opposite directions.
Right: Example of the toy model simulation of
a semi-central collision of two nuclei
with a high string density.
Green arrows  indicate transverse boosts
of the strings with a magnitude above some
threshold.
}
\label{fig:tubes}
\end{figure}

In our study we consider a simplified approach to string-string interaction  mechanism for the case of repulsion. A Monte Carlo (MC) toy model \cite{TOY} is used. 
It is assumed that quark-gluon strings,
formed at early stage of hadron-hadron collision,
may overlap in case of sufficiently high density and interact.
An efficient string-string interaction radius $R_{\rm int}$ is introduced. We consider this free parameter differently from the string radius $r_{0}$. Doubled string-string interaction radius $R_{\rm int}$ can be interpreted as the effective distance of interaction between strings in the transverse plane.
In this MC  model we consider the case of {\it repulsing} strings. We do not take in account neither string attraction nor fusion.

The repulsion mechanism between two strings is considered to be similar to \cite{Abramovsky-1988}: two completely overlapped strings have the energy density of $2E +2E_1$,
while the density of partial overlapping is $2E +2E_1 \cdot S/S_{\rm 0}$. Here $E$ is the energy density of a single free string 
and $E_1$ is the energy density excess due to overlapping. $S$ is the area of the overlap (i.e. it is assumed that effectively interacting strings are "the black discs" with the area $S_{\rm 0}$ in the transverse plane).
Thus, the total energy of the cluster formed by highly overlapped strings, reached in high-energy A-A collision, is larger then the sum of energies of individual separated non-interacting strings.
This energy excess is  
responsible for the string repulsion \cite{Abramovsky-1988}.

In this simplified approach, for any interacting string we consider a coherent sum of  interactions of this string with all strings within the efficient interaction radius.
Thus each string acquires transverse momentum $p_{\rm T}^{\rm string}$ \cite{Abramovsky-1988,TOY},
and all the particles produced during hadronization of this string gain in all region of rapidity %the same
a transverse Lorentz boost. %$\beta_{\rm T}^{\rm string}$
In such a way, due to the string-string interaction, the initial asymmetry of azimuthal configuration of quark-gluon strings could be transferred  into the final state with different harmonics of the azimuthal flow.

 Schematic view of two color flux tubes (strings) boosted apart and generating azimuthally asymmetric flow in the transverse plane due to repulsion is shown
in Figure~\ref{fig:tubes} (left).
An example of the toy model simulation of high string density in the transverse plane in semi-central nucleus-nucleus collision is shown in Figure~\ref{fig:tubes} (right).
One may see that the flow appears as a result of multiple string interactions.

\subsection{Anisotropic flow in the Monte Carlo toy model } 

 We applied the MC model of efficient string repulsion \cite{TOY} for the analysis of two-particle correlation topology in order to study
the origin of the elliptic flow and the higher harmonics observed in nucleus-nucleus collisions at RHIC and LHC.

Two-particle correlation functions are obtained in the MC model
for various centralities of high energy nucleus-nucleus
collisions~\cite{QC-2014-TOY}. They illustrate the onset of collectivity when passing from peripheral to central nucleus-nucleus collisions in a qualitative agreement with RHIC data.  
Different values of $R_{\rm int} =1$ and 2 fm were also tested for evaluation of the contribution of different harmonics and their centrality dependence, see \cite{QC-2014-TOY}.
This qualitative approach demonstrates also the onset of the elliptic flow
and the higher harmonics
in heavy-ion collisions. 

\begin{figure}[t]
\centering
 \includegraphics[width=0.5\textwidth]{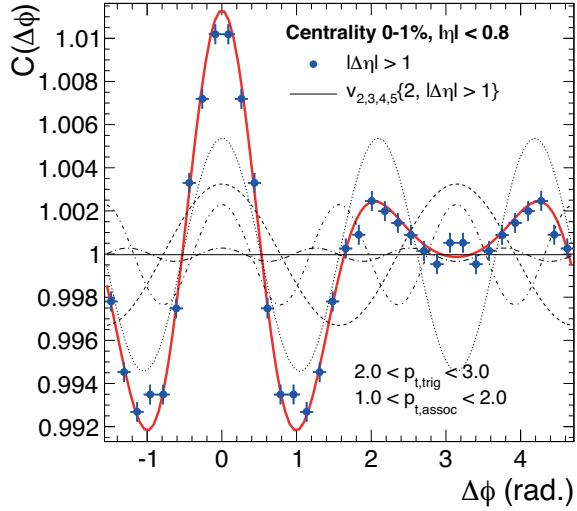}
\caption{The two-particle azimuthal correlation,
measured in $0 < \Delta\phi <\pi$ and shown symmetrized over 2$\pi$, between a trigger particle with $2<\pt< 3$ GeV/$c$ and an associated
particle with $1<\pt < 2$ GeV/$c$ for the 0-1\% centrality
class. The solid red line shows the sum of the measured anisotropic
flow Fourier coefficients v$_2$, v$_3$, v$_4$, and v$_5$ (dashed lines) \cite{ALICE-H}.}
\label{ris:double}
\end{figure}

The measurement of the triangular v${_3}$, quadrangular v${_4}$, and pentagonal v${_5}$ charged particle flow in Pb-Pb collisions
at $\sqrt{s_{NN}}=2.76$ TeV was recently reported in~\cite{ALICE-H}.
In particular, it was shown that one of the remarkable observations, so-called double-ridge structure in very central Pb-Pb collisions
(see Figure~\ref{ris:double}) is related to the triangular flow
and can be understood from the initial spatial anisotropy.

\begin{figure}[h]
%\begin{center}
\centering
$
\begin{array}{ccc}
\begin{overpic}[width=0.5\textwidth, clip=true, trim=10 0 61 0]{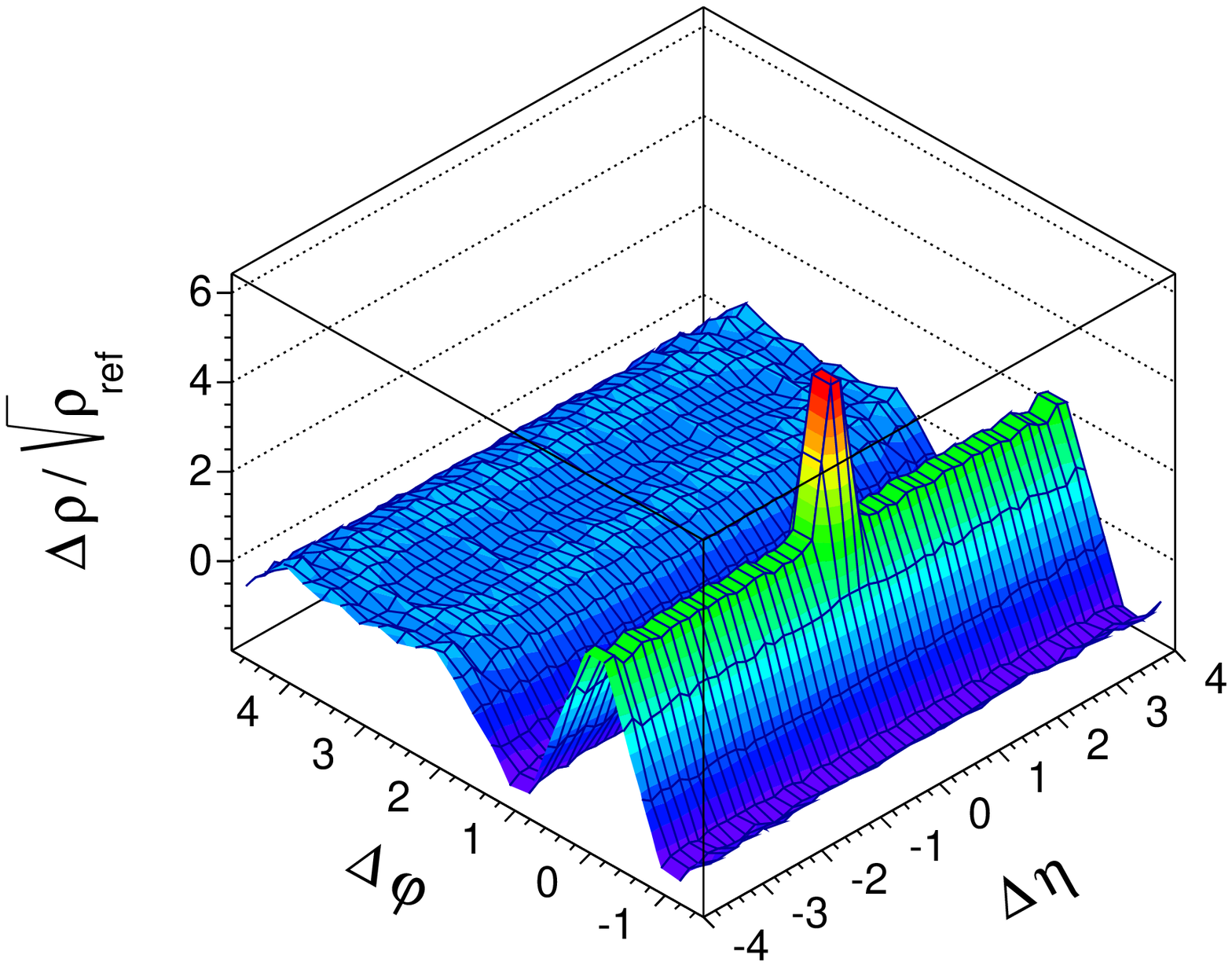}
	\put(0,75){ central events}
	\put(0,70){ $\pt$ 3-5 GeV/c }
\end{overpic}
&
\begin{overpic}[width=0.5\textwidth]{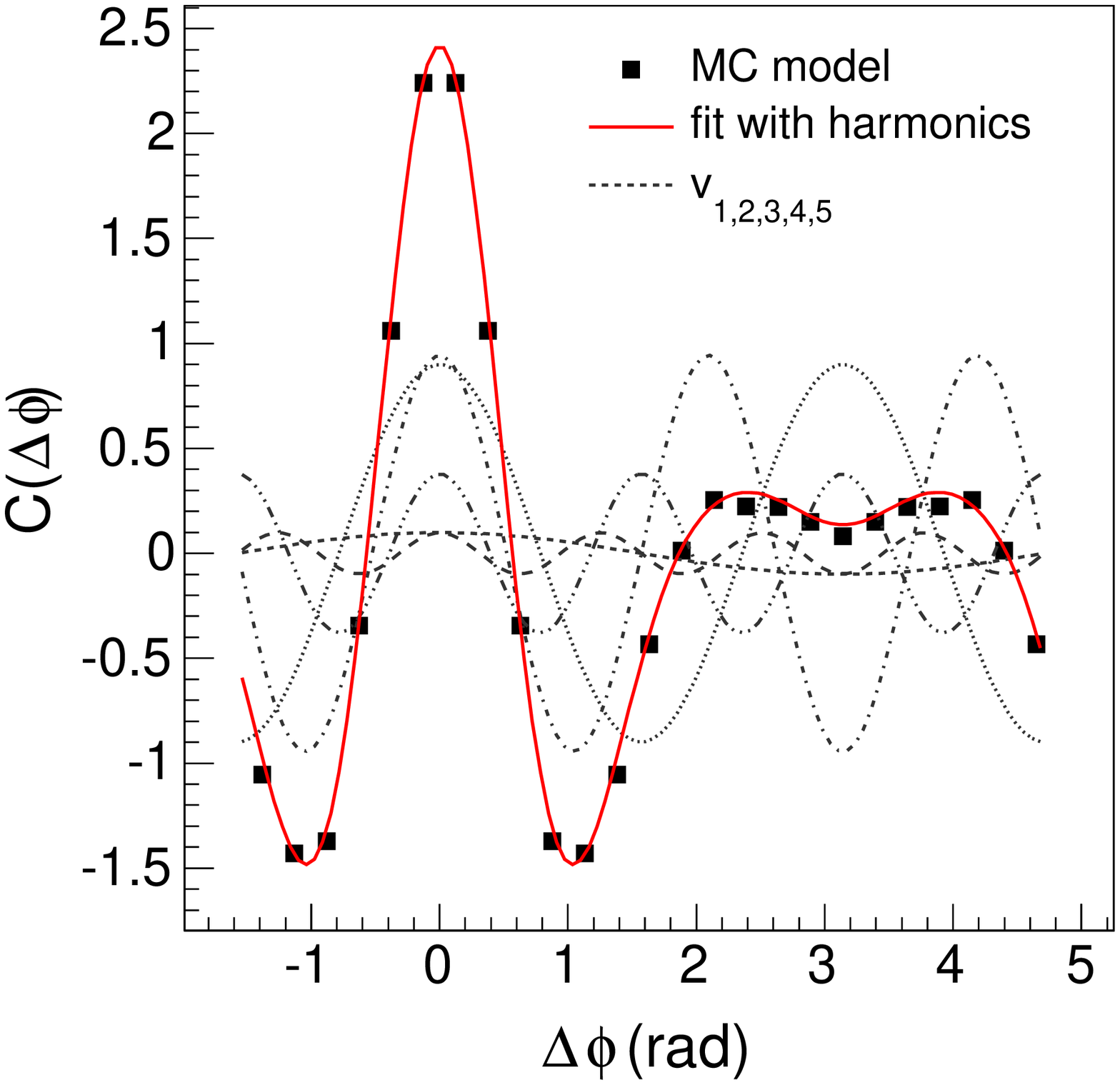}
	\put(50,66){ $\pt$ 3-5 GeV/{\it c} }
\end{overpic}
\end{array}$
%\end{center}
\caption{
Left:
Toy model~\cite{TOY} two-particle correlation function obtained
in the most central events for charged particles with $\pt\in[3,5]$ GeV/{\it c}.
String-string interaction radius $R_{\rm int}$ is 2 fm.
Right:
The harmonic decomposition of the azimuthal profile of the correlation function shown in the left pad.
%same as in (Left) in the transverse plane:
The solid red line shows the sum of
the %obtained
anisotropic
flow Fourier coefficients v$_1$, v$_2$, v$_3$, v$_4$, and v$_5$ (dashed lines).
}
\label{fig:harmonics_dihadron_mostcentral}
\end{figure}

Figure~\ref{fig:harmonics_dihadron_mostcentral} (left) shows
 two-particle azimuthal correlation function obtained in
 the MC model simulations 
of the most central Pb-Pb events, for charged particles with $\pt\in[3,5]$ GeV/{\it c}.
We use the correlation measure $\Delta\rho/\sqrt{\rho_{\rm ref}}$, which is described in detail, in particular, in~\cite{STAR-2012}.
The harmonic decomposition of the azimuthal
profile of this function is presented in
the right pad of Figure~\ref{fig:harmonics_dihadron_mostcentral}.
These data are in a nice %agreement
correspondence with the experimental picture \cite{ALICE-H} shown in Figure~\ref{ris:double} (up to a numerical factor  between the two different observables).

\subsection{Discussion}

All this indicates 
that string percolation %model
with introduced repulsion mechanism between interacting quark-gluon strings, both valence and sea-quark, %is able to reproduce some
may lead to some collective phenomena in nucleus-nucleus collisions. The model gives adequate description of the transition from peripheral to central collisions as well as rise and development of the contribution of the elliptic flow and interplay with higher harmonics.

In these first calculations we neglect the finite string length in rapidity and concentrate on the azimuthal asymmetry of correlation functions. For this case a phenomenological approach of repulsive string-string interaction is shown to be a possible dynamic origin of the observed azimuthal asymmetries of two-particle correlation functions.
Our results show also that the increase of the number of strings and correspondingly the density of the overlapped strings with centrality and collision energy is related mainly to increase of the number of sea-strings.
Therefore, for example, the observed rise with centrality
of the amplitude and a pseudorapidity width
of the so-called same-side 2D Gaussian
obtained in~\cite{Daugherity, STAR-2012} may require an accurate consideration of sea-quark strings formation and their interaction at midrapidity.

The onset of the ridge structure in AA, pA and pp collisions was also considered in the frame of string percolation in a recently published paper~\cite{Pajares-2014}. The increase of the rapidity length  of the effective cluster formed by overlapping sea-strings
is discussed. 
The total energy-momentum of the string cluster is taken here to be the sum of the energy-momenta of the individual strings. 
In our approach, following \cite{Abramovsky-1980, Abramovsky-1988}, the energy of the cluster of the overlapped strings is higher than the sum of individual %single
non-interacting strings, therefore, the effective cluster may be more extended in rapidity.

 Another kind of string interaction, 
string fusion~\cite{Amelin1994,Braun}, 
could explain other observed effects like increased production of strange and multistrange particles  with centrality of nucleus-nucleus collisions. String fusion could be considered as an initial stage leading to glasma or QGP formation.
New constraints on the mentioned models could be obtained from experiment.

\section{Conclusion}

The hypothesis of string-string interaction and percolation string transition looks reasonable in the
quantification of the onset of the low-$\pt$ near side ridge phenomena in Au-Au collisions at RHIC and in Pb-Pb collisions at LHC. One may assume that the onset of string percolation at sufficiently high string densities leads to the formation of rather large clusters composed of overlapped strings extended in rapidity.
Collective effects of interactions between strings inside this cluster could be one of the possible processes leading to repulsion of strings thus shaping topology of two-particle correlation functions. The Monte Carlo toy-model with the efficient account of string repulsion of color flux tubes describes for the first time in a qualitative way the dynamics of the initial conditions of high energy nucleus-nucleus collisions.

The  value of the efficient string-string interaction radius $R_{\rm int}\sim 2$ fm provides qualitative description of elliptic and triangular flows in nucleus-nucleus collisions at RHIC and LHC energies. This radius is found to be larger than the usual string radius, $r_{0}=0.25$ fm.

More detailed quantitative estimates including the case of proton-proton and proton-nucleus collisions will follow.

\acknowledgments
 Authors would like to thank V.~Vechernin for fruitful discussions and for permanent interest to this work. The work was supported for G. F. and I. A. by the St.Petersburg State University grant 11.38.242.2015.

%\endinput
%%
%% End of file `template-8s.tex'.
\end{document}